\documentclass[preprint2,iop,numberedappendix,twocolappendix,appendixfloats]{emulateapj}

\usepackage[caption=false]{subfig}
\usepackage{amsmath}
\usepackage{footnote}
\usepackage{url}
\bibpunct{(}{)}{;}{a}{}{,} 
\newcommand{\dif}{\mathrm{d}}
\newcommand{\bibsortorder}[1]{}

\def\nar{{New~A~Rev.}}          
\def\pasa{{PASA}}               

\shorttitle{Effects of Beam Chromaticity on EoR H{\sc i} Power Spectra Measurements}
\shortauthors{Thyagarajan et~al.}

\def\ASU{\altaffilmark{1}}
\def\ASUtxt{\affil{\altaffilmark{1}Arizona State University, School of Earth and Space Exploration, Tempe, AZ 85287, USA}}

\def\UCB{\altaffilmark{2}}
\def\UCBtxt{\affil{\altaffilmark{2}Dept. of Astronomy, University of California, Berkeley, CA, USA}}

\def\MIT{\altaffilmark{3}}
\def\MITtxt{\affil{\altaffilmark{3}MIT Kavli Institute for Astrophysics and Space Research, Cambridge, MA 02139, USA}}

\def\myemail{\altaffilmark{$\dagger$}}
\def\myemailtxt{\altaffiltext{$\dagger$}{e-mail: t\_nithyanandan@asu.edu}}

\begin{document}

\title{Effects of Antenna Beam Chromaticity on Redshifted 21~cm Power Spectrum and Implications for Hydrogen Epoch of Reionization Array}


\author{
Nithyanandan~Thyagarajan\ASU\myemail,
Aaron~R.~Parsons\UCB,
David~R.~DeBoer\UCB,
Judd~D.~Bowman\ASU,
Aaron~M.~Ewall-Wice\MIT,
Abraham~R.~Neben\MIT, and
Nipanjana~Patra\UCB
}

\ASUtxt
\UCBtxt
\MITtxt
\myemailtxt


\begin{abstract}

Unaccounted for systematics from foregrounds and instruments can severely limit the sensitivity of current experiments from detecting redshifted 21~cm signals from the Epoch of Reionization (EoR). Upcoming experiments are faced with a challenge to deliver more collecting area per antenna element without degrading the data with systematics. This paper and its companions show that dishes are viable for achieving this balance using the Hydrogen Epoch of Reionization Array (HERA) as an example. Here, we specifically identify spectral systematics associated with the antenna power pattern as a significant detriment to all EoR experiments which causes the already bright foreground power to leak well beyond ideal limits and contaminate the otherwise clean EoR signal modes. A primary source of this chromaticity is reflections in the antenna-feed assembly and between structures in neighboring antennas. Using precise foreground simulations taking wide-field effects into account, we provide a framework to set cosmologically-motivated design specifications on these reflections to prevent further EoR signal degradation. We show HERA will not be impeded by such spectral systematics and demonstrate that even in a conservative scenario that does not perform removal of foregrounds, HERA will detect EoR signal in line-of-sight $k$-modes, $k_\parallel \gtrsim 0.2\,h$~Mpc$^{-1}$, with high significance. All baselines in a 19-element HERA layout are capable of detecting EoR over a substantial observing window on the sky.

\end{abstract}
 

\section{Introduction}\label{intro}

The Epoch of Reionization (EoR) is an important period of nonlinear growth of matter density perturbations and astrophysical evolution leading to the large scale structure observed currently in the Universe. The redshifted neutral hydrogen in this epoch has been identified to be one of the most promising and direct probes of the EoR \citep{sun72,sco90,mad97,toz00,ili02}. 

Numerous experiments using low frequency radio telescopes targeting the redshifted 21~cm line from the spin-flip transition of H{\sc i} in this epoch have become operational such as the Murchison Widefield Array \citep[MWA;][]{lon09,bow13,tin13}, the Donald~C.~Backer Precision Array for Probing the Epoch of Reionization \citep[PAPER;][]{par10}, the Low Frequency Array \citep[LOFAR;][]{van13} and the Giant Metrewave Radio Telescope EoR experiment \citep[GMRT;][]{pac13}. Many of these instruments have sufficient sensitivity for a statistical detection of the EoR signal by estimating the spatial power spectrum of the redshifted H{\sc i} spin temperature fluctuations \citep{bea13,thy13}. These instruments are precursors and pathfinders to the next generation of low frequency radio observatories such as the Hydrogen Epoch of Reionization Array\footnote{\url{http://reionization.org/}} \citep[HERA;][]{deb16} and the Square Kilometre Array\footnote{\url{https://www.skatelescope.org/}} (SKA). These next-generation instruments will advance the capability from a statistical detection of the signal to a direct three-dimensional tomographic imaging of H{\sc i} from the EoR. 

The most significant challenge to low frequency EoR observations arises from the extremely bright Galactic and extragalactic foreground synchrotron emission which are $\sim 10^4$ times stronger than the desired EoR signal \citep{dim02,ali08,ber09,ber10,gho12}. However, there are inherent differences in spatial isotropy and spectral smoothness between the EoR signal and the foregrounds \citep[see, e.g.,][]{fur04b,mor04,zal04,san05,fur06,mcq06,mor06,wan06,gle08}. 

When expressed in the coordinate system of power spectrum measurements described by the three-dimensional wavenumber ($k$), the foreground emission is restricted to a wedge-shaped region commonly referred to as the {\it foreground wedge} \citep{bow09,liu09,liu14a,liu14b,dat10,liu11,gho12,mor12,par12b,tro12,ved12,dil13,pob13,thy13,dil14} bounded by the horizon limits \citep{par12b}, whereas the EoR signal has spherical symmetry due to its isotropy. The region of $k$-space excluding the {\it foreground wedge} is commonly referred to as the {\it EoR window} since it is expected to be relatively free of foreground contamination.

Using this knowledge, experiments have begun constraining reionization models \citep{par14,ali15,pob15}. While this progress is encouraging, efforts are slowed by struggles to increase collecting area while contending with foreground and instrumental systematics. The extreme dynamic range required to suppress foreground and instrument systematics demands high precision modeling of foregrounds as observed by modern wide-field instruments. 

Along with companion papers \citep{ewa16,neb16,pat16}, we focus on determining whether larger antenna elements can be viable for EoR experiments. In this paper, we explore the impact of chromaticity in the antenna power pattern on extending foreground power beyond the {\it wedge}. While highlighting the importance of such instrumental systematics on future EoR experiments, we demonstrate that the current HERA instrument design will exceed these specifications to detect the EoR signal with high significance. This will hold true when data is limited by foregrounds and systematics even in the most conservative scenario which involves neither foreground subtraction nor optimal analysis techniques but only relies on a simple spectral weighting scheme. 

This paper is organized as follows. \S\ref{sec:HERA} describes the HERA instrument. A brief summary of the delay spectrum technique used extensively in this analysis and the recently confirmed wide-field instrument effects are presented in \S\ref{sec:delay-spectrum}. \S\ref{sec:sim} describes foreground simulations including antenna beam pattern, all-sky foreground models and two independent EoR models. \S\ref{sec:beam-chromaticity} investigates the effects of chromaticity of antenna beam on the resulting delay power spectrum and the cosmologically motivated constraints it places on reflections in the instrument. EoR-foreground dynamic range of HERA under simple {\it foreground avoidance} criteria are demonstrated in \S\ref{sec:eor-sensitivity}. Our findings are summarized in \S\ref{sec:summary}.

\section{The Hydrogen Epoch of Reionization Array}\label{sec:HERA}

\citet{deb16} describe the HERA instrument in detail while \citet{pob14} explored the range of astrophysical parameters that HERA can probe. We provide a summary of the instrument currently under construction on the South African SKA site.

The primary science goal of HERA is to widen our understanding of the first stars, galaxies, and black holes, and their role in driving reionization. Through power-spectral measurements of the redshifted 21cm line of H{\sc i} in the primordial IGM, HERA aims to directly constrain the topology and evolution of reionization, opening a unique window into the complex astrophysical interplay between the first luminous objects and their environments. HERA builds on the advances of first-generation 21cm EoR experiments, particularly PAPER, MWA, the MIT EoR experiment \citep[MITEoR;][]{zhe14} and the Experiment to Detect the Global EoR Step \citep[EDGES;][]{bow10}.

HERA is deploying 14~m fixed zenith-pointing parabolic dishes that strike an optimal balance between sensitivity and systematics \citep[][and this paper]{ewa16,neb16,pat16}. The large collecting area of a HERA element yields $\approx 5$ times the sensitivity of an MWA tile and more than 20 times that of a PAPER element. 

320 core elements of HERA will be arranged in a compact hexagonal grid, split into three displaced segments to cover the $uv$-plane with sub-element sampling density. The core will be supplemented by 30 additional outrigger elements to tile the $uv$-plane with instantaneously complete sub-aperture sampling out to $250\lambda$ and complete aperture-scale sampling out to $350\lambda$ (at 150~MHz). The layout is discussed in \citet{dil16}.

In this study, we use the HERA-19 hexagonal array layout, a small subset of the planned layout that are currently in use. Without loss of generality, the analysis and results in this paper will apply to the full proposed layout as well.

\section{Delay Spectrum}\label{sec:delay-spectrum}

The delay spectrum technique \citep{par12a,par12b} is briefly described here. We borrow the notation used in \citet{thy15a}. 

{\it Visibilities} measured by an interferometer are given by \citep{van34,zer38,tho01}:
\begin{align}\label{eqn:obsvis}
  V_b(f) &= \int\limits_\textrm{sky} A(\hat{\boldsymbol{s}},f)\,I(\hat{\boldsymbol{s}},f)\,e^{-i2\pi f\frac{\boldsymbol{b}\cdot\hat{\boldsymbol{s}}}{c}}\,\dif\Omega,
\end{align}
where, $\boldsymbol{b}$ is the vector joining antenna pairs (commonly referred to as the baseline vector), $\hat{\boldsymbol{s}}$ is the unit vector denoting direction on the sky, $f$ denotes frequency, $c$ is the speed of light, $\dif\Omega$ is the solid angle element to which $\hat{\boldsymbol{s}}$ is the unit normal vector, $I(\hat{\boldsymbol{s}},f)$ and $A(\hat{\boldsymbol{s}},f)$ are the sky brightness and antenna's directional power pattern, respectively, as a function of $\hat{\boldsymbol{s}}$ and $f$. The {\it delay spectrum}, $\tilde{V}_b(\tau)$, is defined as the inverse Fourier transform of $V_b(f)$ along the frequency coordinate:
\begin{align}\label{eqn:delay-transform}
  \tilde{V}_b(\tau) &\equiv \int V_b(f)\,W(f)\,e^{i2\pi f\tau}\,\dif f,
\end{align}
where, $W(f)$ is a spectral weighting function which can be chosen to control the quality of the delay spectrum \citep{ved12,thy13}, and $\tau$ represents the signal delay between antenna pairs:
\begin{equation}\label{eqn:delay}
  \tau = \frac{\boldsymbol{b}\cdot\hat{\boldsymbol{s}}}{c}.
\end{equation}

The delay spectrum has a close resemblance to cosmological H{\sc i} spatial power spectrum and is defined as:
\begin{align}\label{eqn:delay-power-spectrum}
  P(\boldsymbol{b},k_\parallel) &\equiv |\tilde{V}_b(\tau)|^2\left(\frac{1}{\Omega\Delta B}\right)\left(\frac{D^2\Delta D}{\Delta B}\right)\left(\frac{\lambda^2}{2k_\textrm{B}}\right)^2,
\end{align}
with
\begin{align}
  \boldsymbol{k}_\perp &\equiv \frac{2\pi(\frac{\boldsymbol{b}}{\lambda})}{D}, \label{eqn:kperp-baseline}\\
  k_\parallel &\equiv \frac{2\pi\,f_{21}H_0\,E(z)}{c(1+z)^2}\,\tau, \label{eqn:kprll-delay}
\end{align}
where, $\Delta B$ is the bandwidth, $\lambda$ is the wavelength of the band center, $k_\textrm{B}$ is the Boltzmann constant, $\boldsymbol{k}_\perp$ and $k_\parallel$ are the transverse (on the sky) and line-of-sight (into the sky) wavenumbers respectively, $f_{21}$ is the rest-frame frequency of the 21~cm spin-flip transition of H{\sc i}, $z$ is the redshift, $D\equiv D(z)$ is the transverse comoving distance, $\Delta D$ is the comoving depth along the line of sight corresponding to $\Delta B$, and $h$, $H_0$ and $E(z)\equiv [\Omega_\textrm{M}(1+z)^3+\Omega_\textrm{k}(1+z)^2+\Omega_\Lambda]^{1/2}$ are standard terms in cosmology. $P(\boldsymbol{b},k_\parallel)$ is in units of K$^2 (h^{-1}$~Mpc$)^3$. In this paper, we use $\Omega_\textrm{M}=0.27$, $\Omega_\Lambda=0.73$, $\Omega_\textrm{K}=1-\Omega_\textrm{M}-\Omega_\Lambda$, $H_0=100\,h\,$km$\,$s$^{-1}\,$Mpc$^{-1}$ \citep{wmap9cosmo}. 
\begin{align}
  \Omega\Delta B &= \iint \left|A(\hat{\boldsymbol{s}},f)\,W(f)\right|^2\,\dif\Omega\,\dif f,
\end{align}
is related to the cosmic volume probed by the instrument \citep[see appendix of][]{par14}. 

We note that these visibilities and delay power spectra are dependent on right ascension (RA) of the pointing as well as the frequency band. However, we defer expressing them explicitly as a function of pointing and frequency band (or redshift) to later sections where necessary.

It was recently discovered that in wide-field measurements diffuse foreground emission from wide off-axis angles appears enhanced in the delay spectrum near the edges of the {\it foreground wedge} even on wide antenna spacings \citep{thy15a}. Called the {\it pitchfork} effect, this arises due to severe foreshortening of baseline vectors towards the horizon along joining the antenna pairs thereby enhancing their sensitivity to large scale structures in these directions. Since delay spectrum maps directions on the sky to delay bins, the emission from large scales near the horizon appears enhanced in delay bins near the horizon limits of the {\it foreground wedge}. Since these delay modes lie adjacent to those considered sensitive to the EoR signal, they cause a significant contamination of line-of-sight modes critical for EoR signal detection. These findings were confirmed in MWA observations \citep{thy15b}.

It was also demonstrated in these studies that design of antenna power pattern, specifically its amplitude near the horizon, is an important tool in mitigating foreground contamination caused by these wide-field effects. A dish characterized by a nominal {\it Airy} pattern was found to mitigate this contamination by over four and two orders of magnitude relative to a dipole and a phased array of dipoles respectively. HERA has significantly based its antenna design principles on these findings in choosing its antenna geometry.

In this paper, we investigate the spectral properties of the proposed dish design from a foreground contamination standpoint and the constraints they place on attenuation required to suppress reflections in the instrument. 

\section{Simulations}\label{sec:sim}

We simulate wide-field visibilities for 19-element HERA from all-sky antenna power pattern and foreground models using the PRISim\footnote{The Precision Radio Interferometry Simulator (PRISim) is publicly available at \url{https://github.com/nithyanandan/PRISim}} software package. The simulations cover 24~hr of observation in {\it drift} mode consisting of 80 accumulations spanning 1080~s each. The total bandwidth is 100~MHz centered on 150~MHz consisting of 256 channels with 390.625~kHz frequency resolution. The models used are described below.

\subsection{Antenna Power Pattern}\label{sec:beam-model}

The High Frequency Structural Simulator (HFSS) was used to model the dish and its angular response used in this study. The HFSS model used prime focus optics with a 14~m faceted parabola with a spar f/D ratio of 0.32.  The model has a 1~m central hole in the aluminum surface which is filled with a dielectric material similar to dry soil. The feed used a full PAPER dipole inside of a cylindrical backplane, the backplane is modeled as an aluminum surface. For the metal parts of the dipole, the discs were modeled as aluminum at the actual size, and the arms and terminals were modeled as copper. Dielectric stand-offs and supporting members were included. For the calculations, one pair of arms was excited using a modal port. These models cover a frequency range of 90--210~MHz in intervals of 1~MHz \citep{deb16}. 

For reference, we use two other models for the antenna power pattern. The first is a nominal {\it Airy} pattern corresponding to a uniformly illuminated circular aperture of 14~m diameter and the second is an achromatic model where the response of the design at 150~MHz of the HFSS model described above was fixed as the hypothetical response at all frequencies covering the entire band. This frequency independent model will be used to isolate the effects of spectral structures in the antenna power pattern (or beam chromaticity) on foreground delay power spectra. Hereafter, we refer to these three beams as `simulated', `{\it Airy}' and `achromatic' models.

In a related series of papers, \citet{neb16} discuss the agreement of these simulated antenna beam patterns with actual measurements, \citet{ewa16} model the reflections and return loss expected in the proposed antenna-feed assembly, and \citet{pat16} present measurements quantifying the reflections. Our focus in this paper is to investigate chromaticity of antenna power patterns from the point of view of their impact on foreground contamination.

\subsection{Foreground Model}\label{sec:foreground}

Our all-sky foreground model is the same as the one in \citet{thy15a}. It consists of diffuse emission \citep{deo08} and point sources. The latter is obtained from a combination of the NRAO VLA Sky Survey \citep[NVSS;][]{con98} at 1.4~GHz and the Sydney University Molonglo Sky Survey \citep[SUMSS;][]{boc99,mau03} at 843~MHz with a mean spectral index of -0.83. The diffuse sky model has an angular resolution of 13\farcm 74 and a spectral index estimated for every pixel.

\subsection{EoR Models}\label{sec:EoR-model}

We use two models of EoR. In the first, simulations of the H{\sc i} signal were created using the publicly available 21cmFAST\footnote{\url{http://homepage.sns.it/mesinger/DexM\_\_\_21cmFAST.html}} code described in \citet{mes11}. The code uses the excursion set formalism of \citet{fur04a} to generate ionization and 21cm brightness fields for numerous redshifts. The model shown in this paper assumes the same fiducial values as in \citet{ewa16}: $T_\text{vir}^\text{min} = 2 \times 10^4$~K (virial temperature of minimum mass of dark matter halos that host ionizing sources), $\zeta = 20$ (ionization efficiency), and $R_\text{mfp}=15$~Mpc (mean free path of UV photons) which predicts the redshift of 50\% ionization (and hence a peak in the power spectrum signal) to be at $z=8.5$. The second EoR model is from the simulations described in detail in \citet{lid08}. Hereafter, we refer to these two as EoR models 1 and 2 respectively. 

\section{Chromaticity of Power Pattern}\label{sec:beam-chromaticity}

The equation for delay spectrum defines the interplay between foregrounds and antenna power pattern, and the mapping between geometric phases of baselines and delays. We discuss the spillover of foreground power beyond the horizon delay limits caused by the chromatic nature of the antenna power pattern.

\subsection{Effect on Foreground Contamination}\label{sec:effects-fgdps}

Since our aim in this paper is to quantify EoR-foreground power ratio without employing any sophisticated {\it foreground removal} techniques, we turn our attention to {\it foreground avoidance} strategy that employs spectral weighting technique. Specific choices for spectral weighting function, $W(f)$, have been found to be effective in reducing foreground contamination by many orders of magnitude \citep{thy13} and is regularly used in redshifted 21~cm data analysis \citep{par12a,par12b}. For instance, a {\it Blackman-Harris} function \citep{har78} has a dynamic range of $\sim$100 -- 120~dB in delay power spectrum and a reduced effective bandwidth, with:
\begin{align}\label{eqn:Beff}
  \epsilon &= \frac{1}{B} \int_{-B/2}^{+B/2} |W(f)|^2\,\dif f, \\
  \textrm{and,}\quad B_\textrm{eff} &= \epsilon\,B,
\end{align}
where, $B$ denotes the end-to-end range of the observing band, $1-\epsilon$ is the loss in overall spectral sensitivity and $B_\textrm{eff}$ is the effective bandwidth. The {\it noise equivalent width} is defined as $w_\textrm{ne}\equiv \sum_i |\widetilde{W}(\tau_i)|^2 / |\widetilde{W}(0)|^2$. For a {\it Blackman-Harris} window function denoted by $W_\textrm{BH}(f)$, $\epsilon \approx 50\%$ and $w_\textrm{ne} \approx 2$. 

As the dynamic range required to suppress foreground contamination in the {\it EoR window} may probably be even higher than that provided by a {\it Blackman-Harris} window function, we use a modified version given by convolving a {\it Blackman-Harris} window with itself:
\begin{align}\label{eqn:bhw2}
  W(f) &= W_\textrm{BH}(f) \ast W_\textrm{BH}(f).
\end{align}

\begin{figure}[htb]
  \centering
  \includegraphics[width=\linewidth]{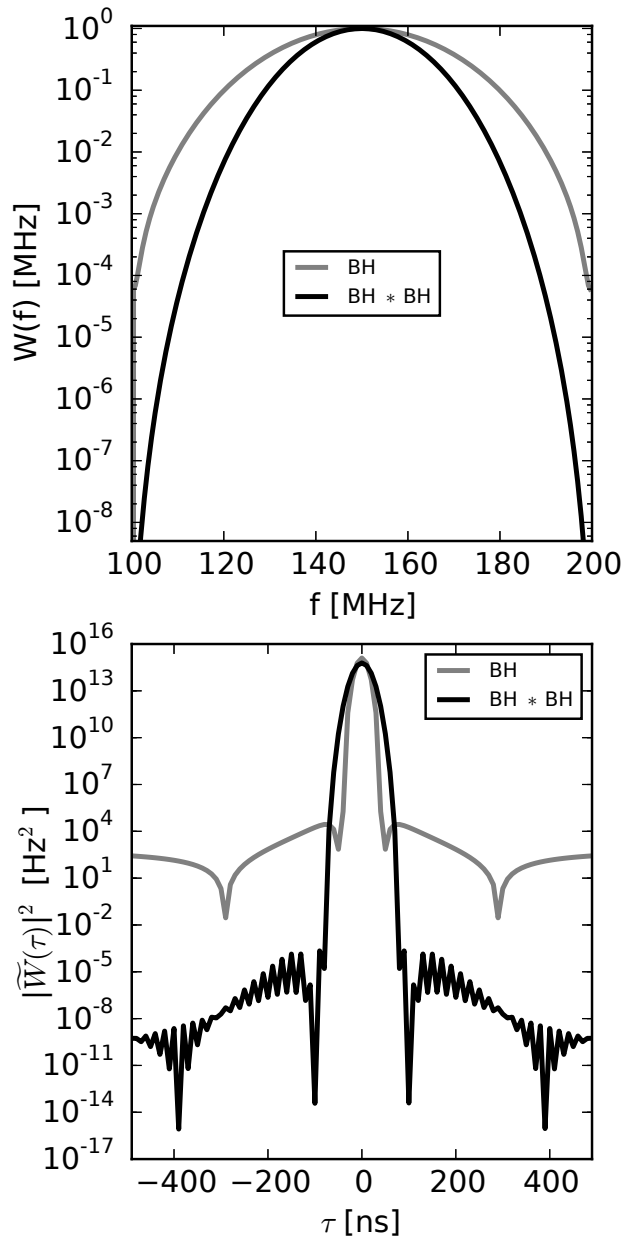}
  \caption{Choices for spectral weighting functions, $W(f)$ ({\it top}) and their delay power spectrum responses, $|\widetilde{W}(\tau)|^2$ ({\it bottom}). The gray curves correspond to a {\it Blackman-Harris} window while those in black correspond to a {\it Blackman-Harris} window convolved with itself. The overall sensitivity of the former is $\epsilon \approx 50\%$ of a rectangular window while that of the latter is $\epsilon \approx 42\%$. As a result, the main lobe of the convolved window function response is slightly wider than that of a {\it Blackman-Harris} window. However, the sidelobes from the convolved window function are suppressed by more than ten orders of magnitude relative to that of a nominal {\it Blackman-Harris} window.}
  \label{fig:window-functions}
\end{figure}

Fig.~\ref{fig:window-functions} shows the {\it Blackman-Harris} window (gray) and  the window in equation~\ref{eqn:bhw2} (black) in frequency (top) and their power responses in delay (bottom). In its Fourier domain, this window has a response obtained by squaring the {\it Blackman-Harris} window response and thus increases the dynamic range further in the power response by another 100 -- 120~dB with $\epsilon\approx 42\%$ and $w_\textrm{ne} \approx 2.88$.

This is an enormous gain in sidelobe characteristics for a small loss of sensitivity relative to a {\it Blackman-Harris} weighting. This will ensure that the contamination resulting from spillover of foregrounds along the line-of-sight $k$-modes are limited only by intrinsic spectral structures in the foregrounds or beam patterns and not by sidelobes from spectral weighting. In this paper, we apply this modified spectral weighting function defined in equation~\ref{eqn:bhw2} whenever foregrounds are represented in the Fourier delay domain. 

\begin{figure*}[htb]
  \centering
  \includegraphics[width=\linewidth]{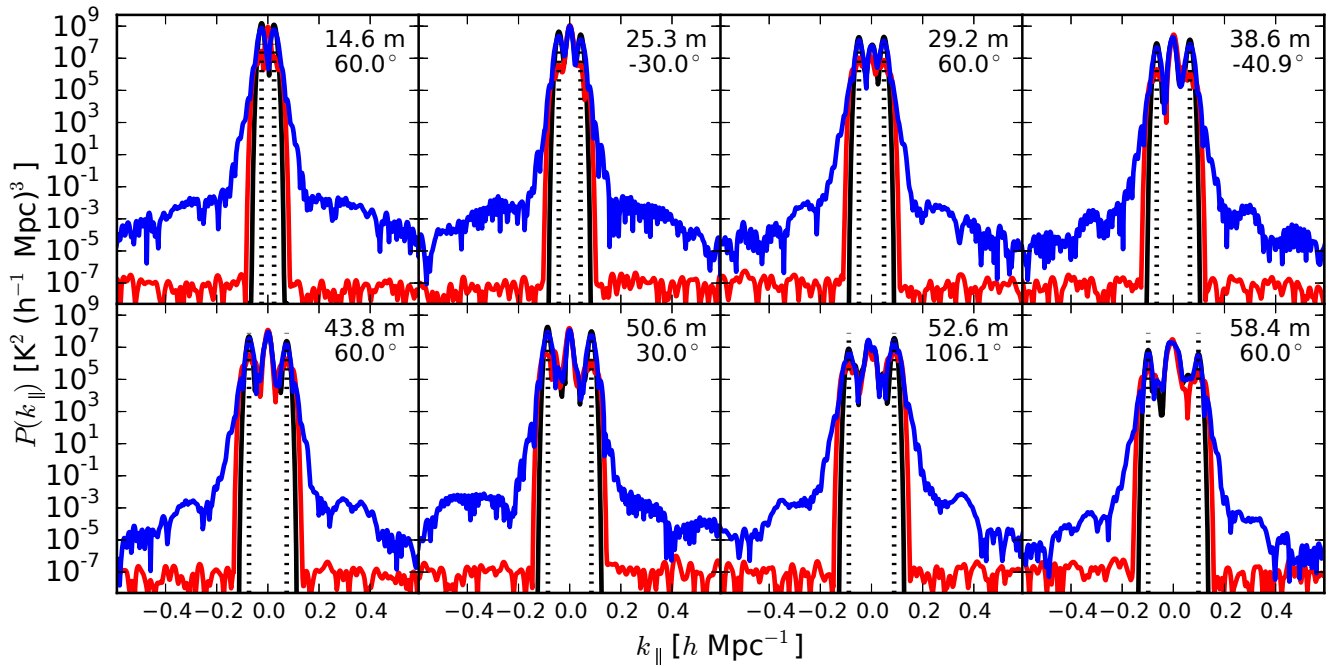}
  \caption{Foreground delay power spectra in units of K$^2 (h^{-1}$~Mpc$)^3$ on unique baseline lengths of HERA-19 at arbitrary sky pointings. The baseline length and orientation (anti-clockwise from East) are annotated at the top right corner of each panel. Black, red and blue curves correspond to delay power spectra obtained with achromatic, {\it Airy} and chromatic simulated antenna beams respectively. The achromatic beam has no spectral structure, the simulated chromatic beam has maximum chromaticity while a nominal {\it Airy} pattern has intermediate level of chromaticity. The {\it pitchfork} effect is clearly visible as peaks around the horizon limits. With increase in chromaticity of the antenna beam the foreground spillover beyond the {\it foreground wedge} becomes progressively worse - the extent of foreground-spillover wings beyond the horizon limits (vertical dotted lines) and the amplitude of spillover beyond $k_\parallel\gtrsim 0.2\,h$~Mpc$^{-1}$ is most severe for the chromatic simulated beam, intermediate for an {\it Airy} pattern and negligible for an achromatic beam.}
  \label{fig:asm-dps-beam-chromaticity-baselines}
\end{figure*}

In HERA-19 array layout, there are 30 unique baseline vectors and 8 unique baseline lengths. Fig.~\ref{fig:asm-dps-beam-chromaticity-baselines} shows the delay power spectra of foregrounds on the 8 unique baseline lengths obtained with the aforementioned models for power pattern at arbitrary pointings on the sky. In all these panels, the full-band foreground delay power spectra obtained with achromatic, {\it Airy} and chromatic simulated beam patterns are shown in black, red, and blue respectively. The brightening of foreground power near the horizon limits (vertical dotted lines) due to the {\it pitchfork} effect \citep{thy15a,thy15b} is prominently seen in all cases. A clear broadening of spillover-wings outside the horizon limits is seen with increasing chromaticity as the beam is changed from the achromatic to the {\it Airy} to the chromatic simulated model. For instance, the spillover from foreground delay power spectrum obtained with chromatic beam pattern is restricted to $|k_\parallel| \lesssim 0.2\,h$~Mpc$^{-1}$, with the {\it Airy} pattern it is restricted to $|k_\parallel| \lesssim 0.15\,h$~Mpc$^{-1}$, while with the achromatic beam it is restricted to $|k_\parallel| \lesssim 0.12\,h$~Mpc$^{-1}$ even on longest baseline lengths. It is also noted that despite the extreme dynamic range of the spectral weighting function employed, the tail of the foreground spillover at $|k_\parallel| \gtrsim 0.2\,h$~Mpc$^{-1}$ with an {\it Airy} beam pattern is many orders of magnitude higher than that using an achromatic beam while that from the simulated chromatic beam is even higher than from an {\it Airy} pattern by a few orders of magnitude. Thus, the foreground spillover shown is truly limited by intrinsic spectral features in the antenna beam patterns. 

This clearly demonstrates that with increasing beam chromaticity, the foreground contamination inherently extends farther along $k_\parallel$. Thus, the chromaticity of antenna beam needs to be controlled in EoR experiments to keep the foreground systematics sufficiently low.

Such a significant spillover is not bound by the horizon limits as this is caused by spectral structure in the antenna beam pattern and is independent of geometric phases. Hence, delay-based complex deconvolution techniques \citep{par09,par12b} that rely on smoothness of foreground spectra and only spectral window shape will not have adequate information to accurately deconvolve intrinsic supra-horizon spillover arising from the chromaticity of the antenna beam. While delay power spectrum estimation using {\it foreground removal} strategy that accounts for direction- and beam chromaticity-dependent effects is possible, we leave it for future work.

\subsection{Constraints on Reflections in the Instrument}\label{sec:constraints-reflectometry}

A primary cause for spectral structure in antenna power patterns is reflections in the instrument. \citet{pat16} and \citet{ewa16} discuss the measured and simulated reflections respectively between a dish and its feed for HERA. Reflections between different antennas also causes chromaticity in the antenna beam. In this section, we provide cosmologically motivated design specifications on instrument systematics caused by these two types of reflections. 

Reflections shift the measured foreground power to higher modes in $\tau$ (or in $k_\parallel$) and thus cause further contamination in these critical higher $k_\parallel$-modes. These delay shifts introduce ripples in the spectrum. The net chromaticity in the measurements is the product of spectral structure arising out of the intrinsic nature of foreground emission, $I(\hat{\boldsymbol{s}},f)$, the baseline- and position-dependent frequency structure of the geometric phases, $e^{-i2\pi f\frac{\boldsymbol{b}\cdot\hat{\boldsymbol{s}}}{c}}$, the spectral features in the antenna power pattern, $A(\hat{\boldsymbol{s}},f)$, besides any other spectral structures in the instrument unaccounted for. In $P(\boldsymbol{b},k_\parallel)$ defined in the Fourier domain, these factors have a convolving effect. 

From these factors, we isolate here the effects of chromaticity in the antenna power pattern caused by reflections in antenna structures and signal paths. And, we devise a cosmologically-motivated method to set design requirements on suppressing reflections in the instrument. 

We define the required attenuation on the reflected foreground power as the ratio: 
\begin{align}\label{eqn:attenuation}
  \Gamma_{k_\textrm{p}}(\boldsymbol{b},\tau) \geq \max_{|k_\parallel|>k_\textrm{p}}\left\{\frac{\left\langle P_\textrm{FG}(\boldsymbol{b},k_\parallel - \frac{\dif k_\parallel}{\dif \tau}\,\tau)\right\rangle}{P_\textrm{H{\sc i}}(|\boldsymbol{b}|,k_\parallel)}\right\}^{1/2},
\end{align}
where, $\tau$ is the delay caused by reflections, and $\dif k_\parallel/\dif \tau$ is the {\it jacobian} in the transformation of $\tau$ to $k_\parallel$ (see equation~\ref{eqn:kprll-delay}). $P_\textrm{H{\sc i}}(|\boldsymbol{b}|,k_\parallel)$ is the EoR H{\sc i} delay power spectrum. $P_\textrm{FG}(\boldsymbol{b},k_\parallel)$ is the full-band foreground delay power spectrum. The angular brackets denote averaging over baseline vectors of the specified length, and a local sidereal time (LST) range. 

We adopt the interpretation that reflections cause additional spectral structures in the antenna power pattern in which they would be absent otherwise. Thus, we use an antenna power pattern in which those spectral structures from reflections which we are interested in addressing are absent but the intrinsic chromaticity of foregrounds and the geometric phases of baselines are included. $\Gamma_{k_\textrm{p}}(\tau)$ is determined by the requirement that the reflected foreground power so obtained, after shifting in delay, lies below the EoR H{\sc i} signal power in line-of-sight spatial scales of interest, $|k_\parallel|>k_\textrm{p}$. Thus, by disentangling spectral systematics inherent in the antenna power pattern from the overall EoR-foreground dynamic range, a design requirement that sets a limit on that systematic is obtained.

Both EoR models were employed in this analysis to reduce the dependence of design specifications on any single model. However, we found similar results except for minor differences caused by small amplitude differences between the two EoR models. Hence, we only present results from using EoR model 1 at 150~MHz.

\subsubsection{Reflections in Dish-Feed Assembly}\label{sec:dish-feed-reflections}

\begin{figure}[htb]
  \centering
  \includegraphics[width=\linewidth]{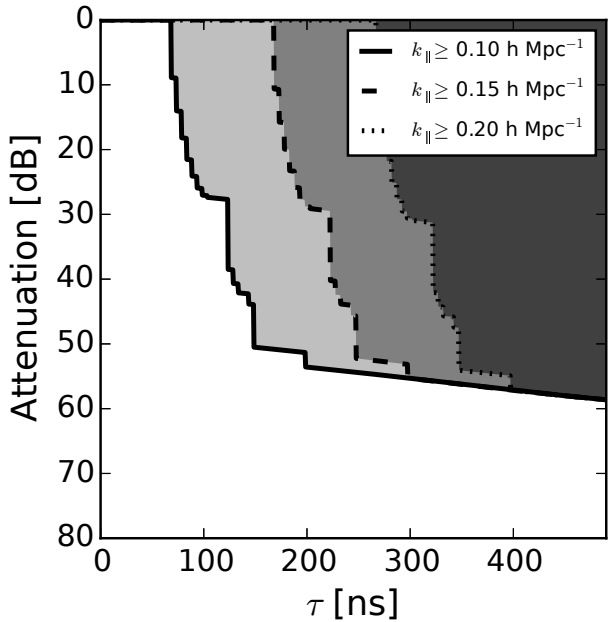}
  \caption{Minimum required attenuation (in dB) for internal dish-feed reflections required to keep the reflected foreground power below EoR H{\sc i} signal power for all $k_\parallel$-modes greater than $0.1\,h$~Mpc$^{-1}$ (solid), $0.15\,h$~Mpc$^{-1}$ (dashed) and $0.2\,h$~Mpc$^{-1}$ (dotted). This is obtained on 14.6~m antenna spacing for EoR model 1 at 150~MHz ($z\approx 8.47$) and foreground power obtained with an achromatic antenna beam model. For EoR to be detectable in respective $k_\parallel$-modes despite these systematics, the attenuation of reflections must exceed these limits (outside the shaded regions).}
  \label{fig:fg-reflections-achrmbeam}
\end{figure}

For reflections caused by dish-feed assembly, we use $P_\textrm{FG}(k_\parallel)$ obtained with the achromatic beam model on 14.6~m baselines in a 0--12 hr LST range. In this instance, we assume dish-feed reflections predominantly imprint chromaticity in an otherwise achromatic antenna power pattern and $\Gamma_{k_\textrm{p}}(\tau)$ is interpreted as the design goal for attenuating these reflections. These reflections are primarily caused in the dish-feed assembly and to some extent in other subsystems not accounted for. Fig.~\ref{fig:fg-reflections-achrmbeam} shows $\Gamma_{k_\textrm{p}}(\tau)$ (in dB) for $k_\textrm{p}$ chosen to be 0.1~$h$~Mpc$^{-1}$ (solid), 0.15~$h$~Mpc$^{-1}$ (dashed), and 0.2~$h$~Mpc$^{-1}$ (dotted). These curves set an upper limit for the reflected foreground power to lie below the EoR H{\sc i} signal power as a function of delays. It implies that if attenuated to levels that lie in the regions below the different shaded regions, such spectral systematics in the instrument will not hinder detection of EoR in those corresponding $k_\parallel$-modes of interest. 

The elbow-shaped turnover is a measure of the most severe requirement on attenuation of reflections. This depends sensitively on the choice of $k_p$. For instance, the attenuation required is $\gtrsim 54$~dB at $\sim 200$~ns for $k_\textrm{p}=0.1\,h$~Mpc$^{-1}$, $\gtrsim 56$~dB at $\sim 300$~ns for $k_\textrm{p}=0.15\,h$~Mpc$^{-1}$ and $\gtrsim 58$~dB at $\sim 400$~ns for $k_\textrm{p}=0.2\,h$~Mpc$^{-1}$. Measurements are underway \citep{pat16} to confirm that the return losses in the HERA design lie within the regions excluded by the shaded regions.

\subsubsection{Antenna-to-Antenna Reflections}\label{sec:antenna-antenna-reflections}

\begin{figure*}[htb]
  \centering
  \includegraphics[width=\linewidth]{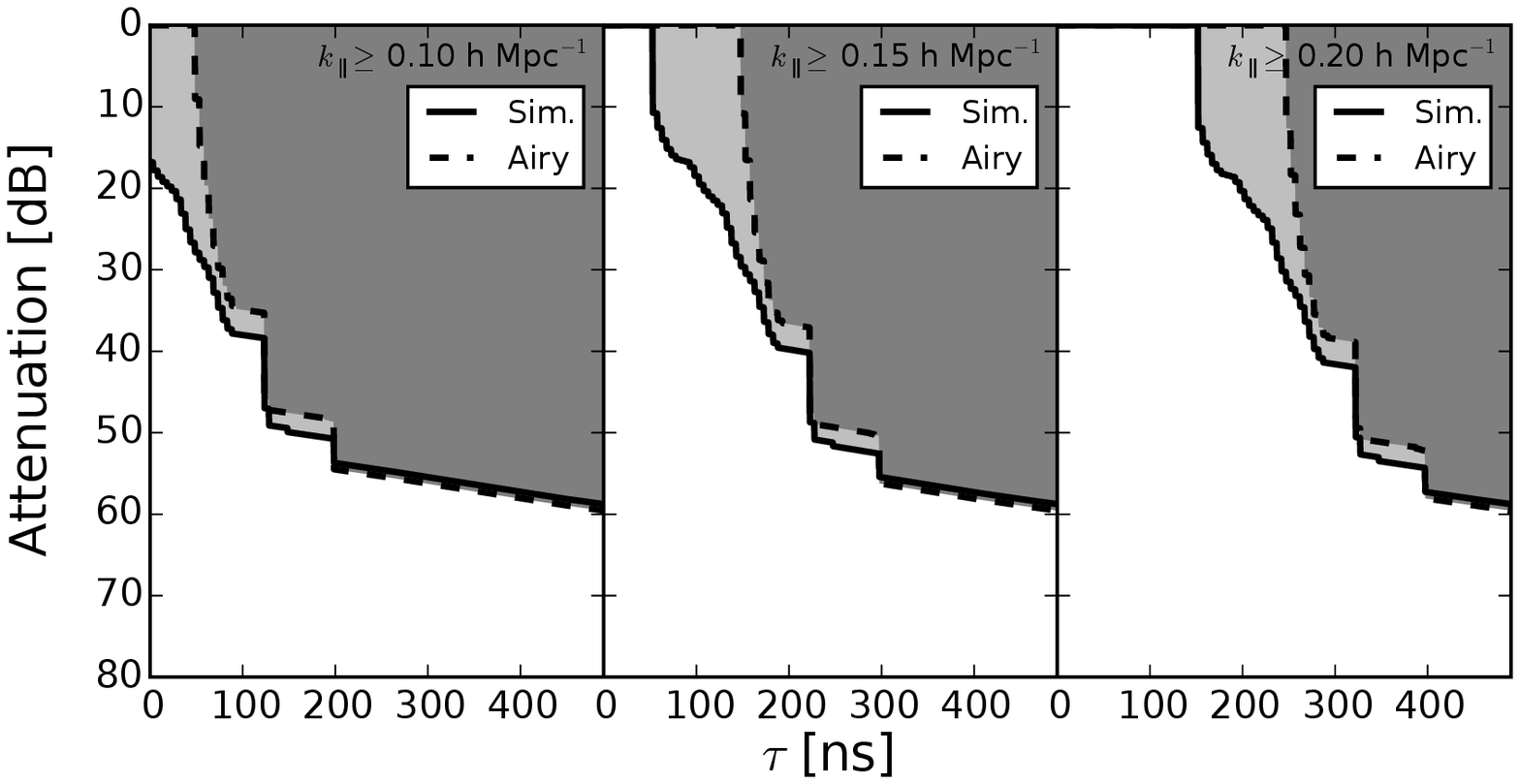}
  \caption{Minimum required attenuation (in dB) for antenna-to-antenna reflections to keep reflected foreground power below EoR H{\sc i} signal power for all $k_\parallel$-modes greater than $0.1\,h$~Mpc$^{-1}$ (left), $0.15\,h$~Mpc$^{-1}$ (middle) and $0.2\,h$~Mpc$^{-1}$ (right). This is obtained on 14.6~m antenna spacing for EoR model 1 at 150~MHz ($z\approx 8.47$) and foreground power obtained with an {\it Airy} (dashed) and simulated chromatic (solid) antenna beam models. Regions excluding shaded regions indicate EoR will be detectable in respective $k_\parallel$-modes despite these reflections. Increase in beam chromaticity makes the requirement on attenuation more severe. For instance, $k_\parallel > 0.1\,h$~Mpc$^{-1}$ modes will be inaccessible with the simulated chromatic beam if these reflections are taken into account. However, for the same beam the instrument will have much higher tolerance despite these reflections if the modes of interest are $k_\parallel > 0.2\,h$~Mpc$^{-1}$. Due to lower chromaticity, an {\it Airy} beam offers more tolerance to reflections than the simulated chromatic beam.}
\label{fig:fg-reflections}
\end{figure*}

Similarly, for reflections arising out of structures and interfaces across different antennas, we use $P_\textrm{FG}(k_\parallel)$ obtained with the simulated chromatic and {\it Airy} antenna beams on 14.6~m baselines. The simulated chromatic pattern in fact includes explicit simulations of reflections between structures within an antenna subsystem. We assume the same is true for the {\it Airy} pattern as well. But both these beams do not include spectral features from antenna-to-antenna reflections. Thus $\Gamma_{k_\textrm{p}}(\tau)$ provides lower limit for attenuation of such antenna-to-antenna reflections. Fig.~\ref{fig:fg-reflections} is similar to Fig.~\ref{fig:fg-reflections-achrmbeam} with constraints for values of $k_\textrm{p}$ chosen to be 0.1~$h$~Mpc$^{-1}$ (left), 0.15~$h$~Mpc$^{-1}$ (middle), and 0.2~$h$~Mpc$^{-1}$ (right). Lower limits are estimated for {\it Airy} (dashed lines) and simulated chromatic beams (solid lines).

Increase in beam chromaticity makes the lower limits on attenuating reflections more severe relative to that from an achromatic beam. For instance, for $k_\textrm{p}=0.1\,h$~Mpc$^{-1}$ there is no delay at which the required attenuation is unconstrained to the left of the elbow-shaped turnover, including at $\tau=0$. This means that when additional chromaticity from antenna-to-antenna reflections is taken into account, the requirement that all $k_\parallel$-modes with $k_\parallel > 0.1\,h$~Mpc$^{-1}$ be accessible for EoR signal detection will not be satisfied with the currently simulated chromatic beam for HERA. For $k_\textrm{p}=0.15\,h$~Mpc$^{-1}$, there is a narrow range of allowed attenuation which is unconstrained for $\tau \lesssim 50$~ns. However, for $k_\parallel > 0.2\,h$~Mpc$^{-1}$, the foreground power spectrum using simulated chromatic dish beam will have enough room to be effective for EoR signal detection despite additional chromaticity arising from antenna-to-antenna reflections. 

We note here that designing a dish whose antenna beam closely resembles a nominal {\it Airy} pattern will have a significantly higher tolerance for allowing antenna-to-antenna reflections and yet remain very effective for EoR signal detection. The HERA collaboration constantly pursues improvement of its dish design to minimize limitations from such chromatic systematics.

Similar design limits on reflections on various antenna spacings were also studied. With increase in baseline length, the amplitude of foreground power decreases but the {\it foreground window} also widens. This means the amplitude of reflections required to be suppressed is lower but there is lesser room along delay axis before the reflected foreground modes shift to and contaminate the $k_\parallel$-modes of interest. As a result, the elbow-shaped design constraints shift upward and leftward. For instance, in the presence of dish-feed reflections, it will still be feasible to probe only $k_\parallel > 0.15\,h$~Mpc$^{-1}$ with the 58.4~m baselines.

\section{EoR-Foreground Dynamic Range}\label{sec:eor-sensitivity}

We estimate the EoR-foreground dynamic range (ratio of EoR signal and foreground powers) for both EoR models in the presence of antenna beam chromaticity, using HERA as an example. In order to avoid signal evolution across the entire band, we use subbands in which EoR signal evolution is not expected to be significant. These subbands with $B_\textrm{eff}=10$~MHz are defined by spectral weights in equation~\ref{eqn:bhw2}. These subbands are centered at 150~MHz and 170~MHz to estimate EoR-foreground dynamic range at $z \approx 8.47$ and $z \approx 7.36$ respectively.

\subsection{Dependence on Baseline-$k_\parallel$ Parameter Space}\label{sec:baseline-kprll}

\begin{figure*}[htb]
  \centering
  \subfloat[][150~MHz subband ($z \approx 8.47$)]{\label{fig:asm-dps-beam-chromaticity-baselines-150-subband}\includegraphics[width=\linewidth]{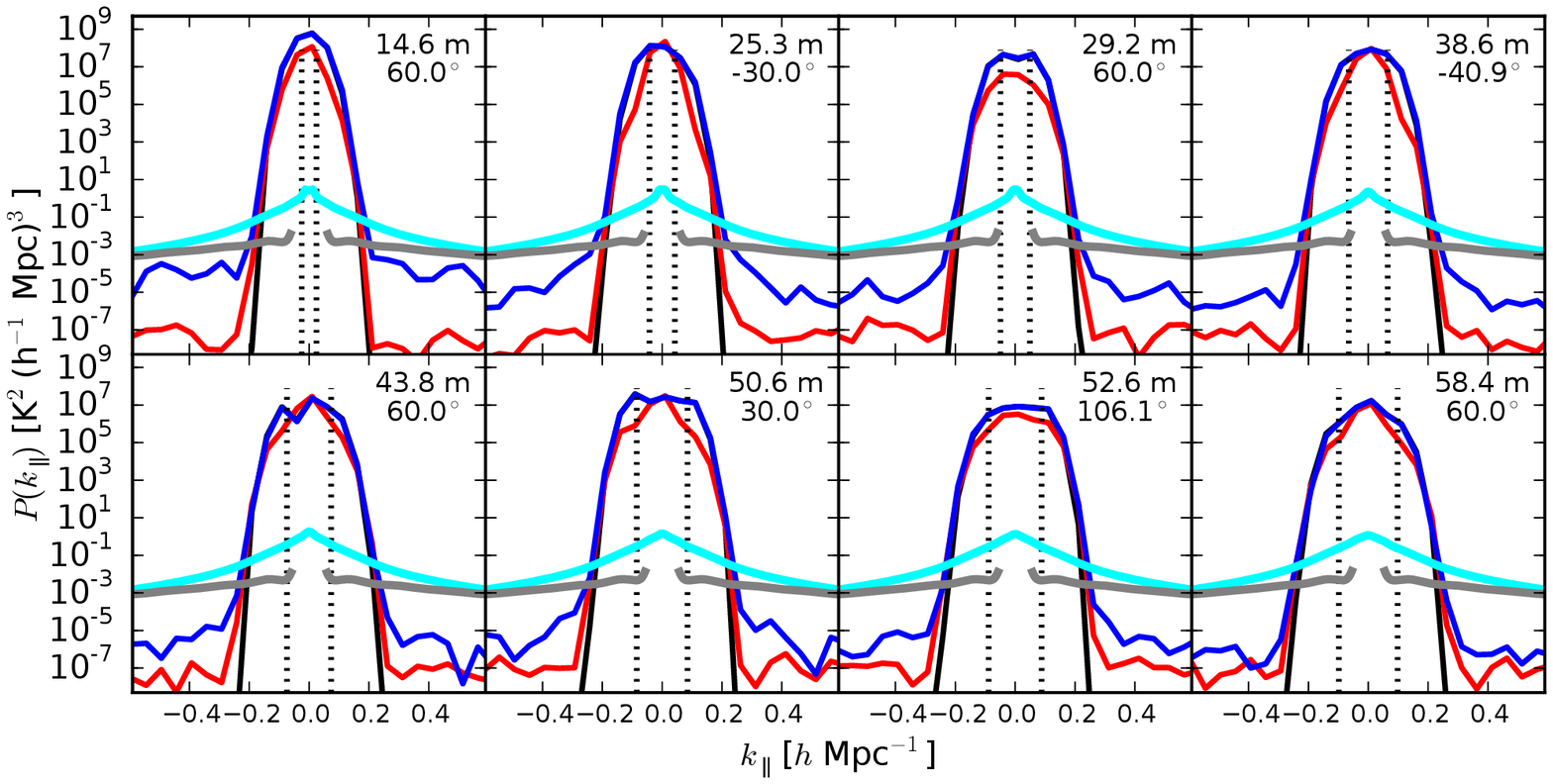}} \\
  \subfloat[][170~MHz subband ($z \approx 7.36$)]{\label{fig:asm-dps-beam-chromaticity-baselines-170-subband}\includegraphics[width=\linewidth]{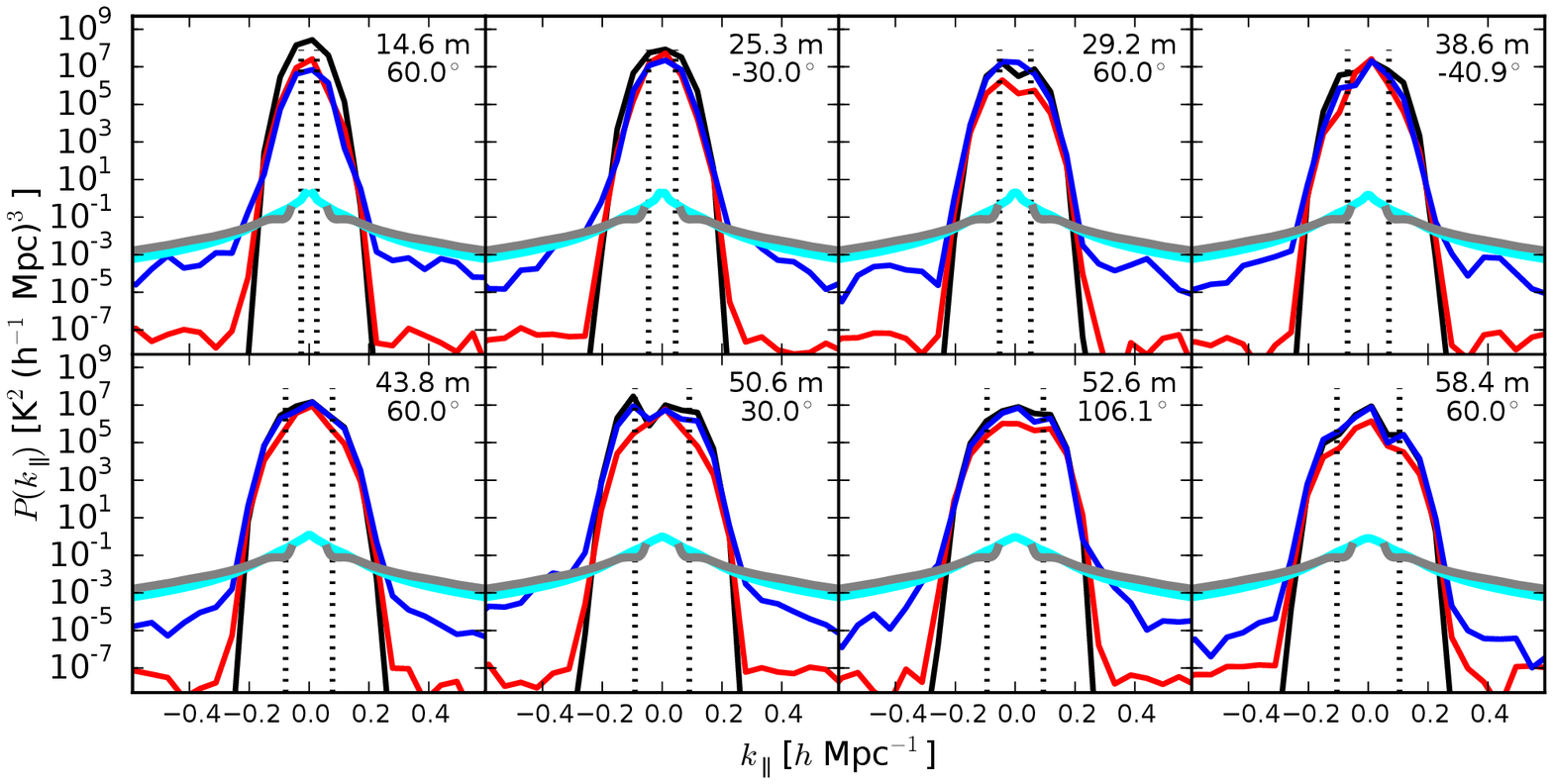}}
  \caption{EoR signal and foreground delay power spectrum in units of K$^2 (h^{-1}$~Mpc$)^3$ in 150~MHz (top) and 170~MHz (bottom) subbands ($B_\textrm{eff}=10$~MHz) on eight unique baseline lengths of HERA-19 at arbitrary sky pointings. The length and orientation of the baseline vector corresponding to each panel is annotated on the top right corner. EoR models 1 and 2 are shown in cyan and gray respectively. The foregrounds obtained with achromatic, {\it Airy} and simulated chromatic antenna beams are shown in black, red and blue respectively. EoR-foreground dynamic range is highest for antenna beam with least chromaticity and vice versa. Even for the simulated chromatic antenna beam pattern, which has the highest chromaticity among the antenna beam models considered, foreground power will be lower than signal power from the two independent EoR models by more than two orders of magnitude for $|k_\parallel| \gtrsim 0.2\,h$~Mpc$^{-1}$ on all HERA baselines.}
  \label{fig:subbands}
\end{figure*}

We estimate EoR-foreground dynamic range for HERA using antenna beams of different chromaticities. Fig.~\ref{fig:asm-dps-beam-chromaticity-baselines-150-subband} and \ref{fig:asm-dps-beam-chromaticity-baselines-170-subband} show the EoR signal and foreground power in 150~MHz and 170~MHz subbands respectively. EoR models 1 and 2 are shown in cyan and gray respectively. The foreground delay power spectra obtained with achromatic, {\it Airy} and simulated chromatic antenna patterns are shown in black, red, and blue respectively. Each panel corresponds to a baseline vector, same as in Fig.~\ref{fig:asm-dps-beam-chromaticity-baselines}. 

Due to narrower subbands, the resolution of the foreground delay spectrum is coarser and the central region of foreground contamination extends to $|k_\parallel| \lesssim 0.2\,h$~Mpc$^{-1}$. The coarsening of delay resolution significantly absorbs the distinct differences caused by beams of different chromaticities as seen in the full-band foreground delay spectra outside the {\it foreground wedge} in Fig.~\ref{fig:asm-dps-beam-chromaticity-baselines}. This means the foreground spillover wings that extend beyond the horizon even in the case of simulated chromatic beam are on spectral scales larger than the 10~MHz effective bandwidth. Hence, this results in an EoR signal-foreground crossover at $|k_\parallel| \gtrsim 0.2\,h$~Mpc$^{-1}$ independent of different antenna beam chromaticities used in this analysis. 

From Fig.~\ref{fig:subbands}, it is clearly demonstrated that across all baselines and subbands, HERA should detect EoR by more than two orders of magnitude above foreground contamination obtained with any level of antenna beam chromaticity. 

\subsection{Dependence on Baseline-RA Parameter Space}\label{sec:baseline-RA}

We investigate the EoR-foreground dynamic range in two dimensional parameter space formed by baselines and pointings in Right Ascension (RA$\equiv\alpha$) to highlight capabilities of HERA and provide clues for the best observing window. 

Due to the assumed isotropy of the redshifted H{\sc i} power spectrum, it is independent of $\alpha$. However, it is dependent on the center of the subband, $f_0$, where $f_0 = f_{21}/(1+z)$. The foreground delay power spectra depends on both $f_0$ and $\alpha$. Thus, we rewrite the EoR H{\sc i} and foreground delay power spectra explicitly as a function of these quantities as $P_\textrm{H{\sc i}}(|\boldsymbol{b}|,k_\parallel,f_0)$ and $P_\textrm{FG}(\boldsymbol{b},k_\parallel,f_0,\alpha)$ respectively.

We define baseline-dependent $k_\parallel$-modes of interest for {\it foreground avoidance} as $|k_\parallel| > k_\parallel^\textrm{FA}(\boldsymbol{b})$, with:
\begin{align}
  k_\parallel^\textrm{FA}(\boldsymbol{b}) &= \frac{2\pi\,f_{21}H_0\,E(z)}{c(1+z)^2}\,\left(\frac{|\boldsymbol{b}|}{c} + \frac{\zeta}{B_\textrm{eff}}\right), \label{eqn:kprll-limit}
\end{align}
where, the first term inside the parenthesis denotes the horizon limit, the second relates to the resolution due to subband bandwidth, $B_\textrm{eff}$, and $\zeta$ denotes a buffer to ensure the $k_\parallel^\textrm{FA}(|\boldsymbol{b}|)$ threshold ``safely'' avoids the main lobe of foreground power. In this study, we use $\zeta=3.5$.

We estimate the worst case EoR-foreground dynamic range in the $\boldsymbol{b}$--$\alpha$ parameter space as:
\begin{align}
  \rho(\boldsymbol{b},f_0,\alpha) &= \min_{|k_\parallel|>k_\parallel^\textrm{FA}(\boldsymbol{b})}\left\{\frac{P_\textrm{H{\sc i}}(|\boldsymbol{b}|,k_\parallel,f_0)}{P_\textrm{FG}(\boldsymbol{b},k_\parallel,f_0,\alpha)}\right\}.
\end{align}
This denotes the minimum EoR-foreground dynamic range in $k_\parallel$-modes of interest for {\it foreground avoidance}, $|k_\parallel| > k_\parallel^\textrm{FA}(\boldsymbol{b})$, for each baseline vector ($\boldsymbol{b}$), subband ($f_0$), and sky pointing ($\alpha$).

\begin{figure*}[htb]
  \centering
  \subfloat[][EoR model 1, 150~MHz subband ]{\label{fig:21cmfast-fg-ratio-150}\includegraphics[width=0.45\linewidth]{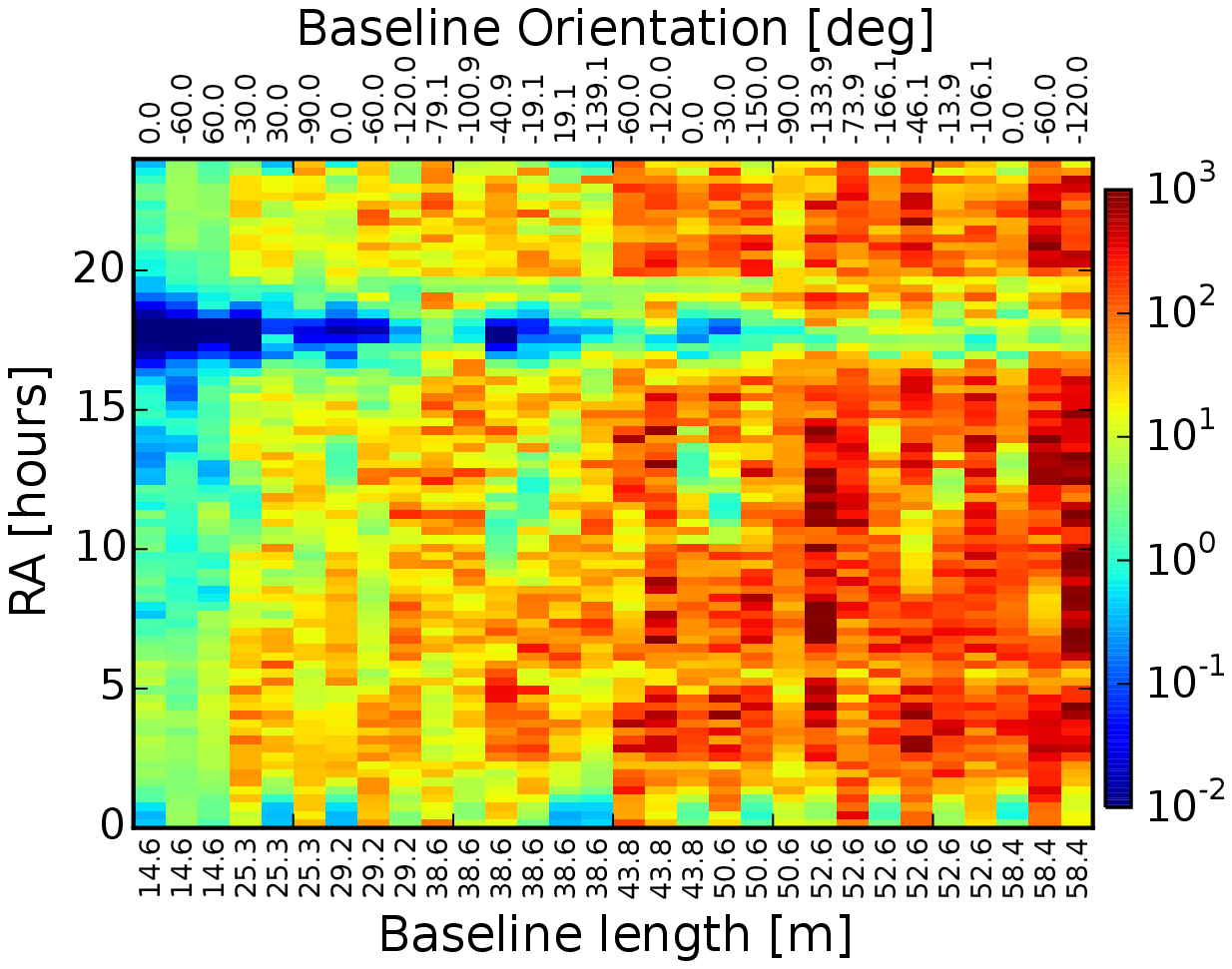}}
  \subfloat[][EoR model 2, 150~MHz subband ]{\label{fig:lidz-fg-ratio-150}\includegraphics[width=0.45\linewidth]{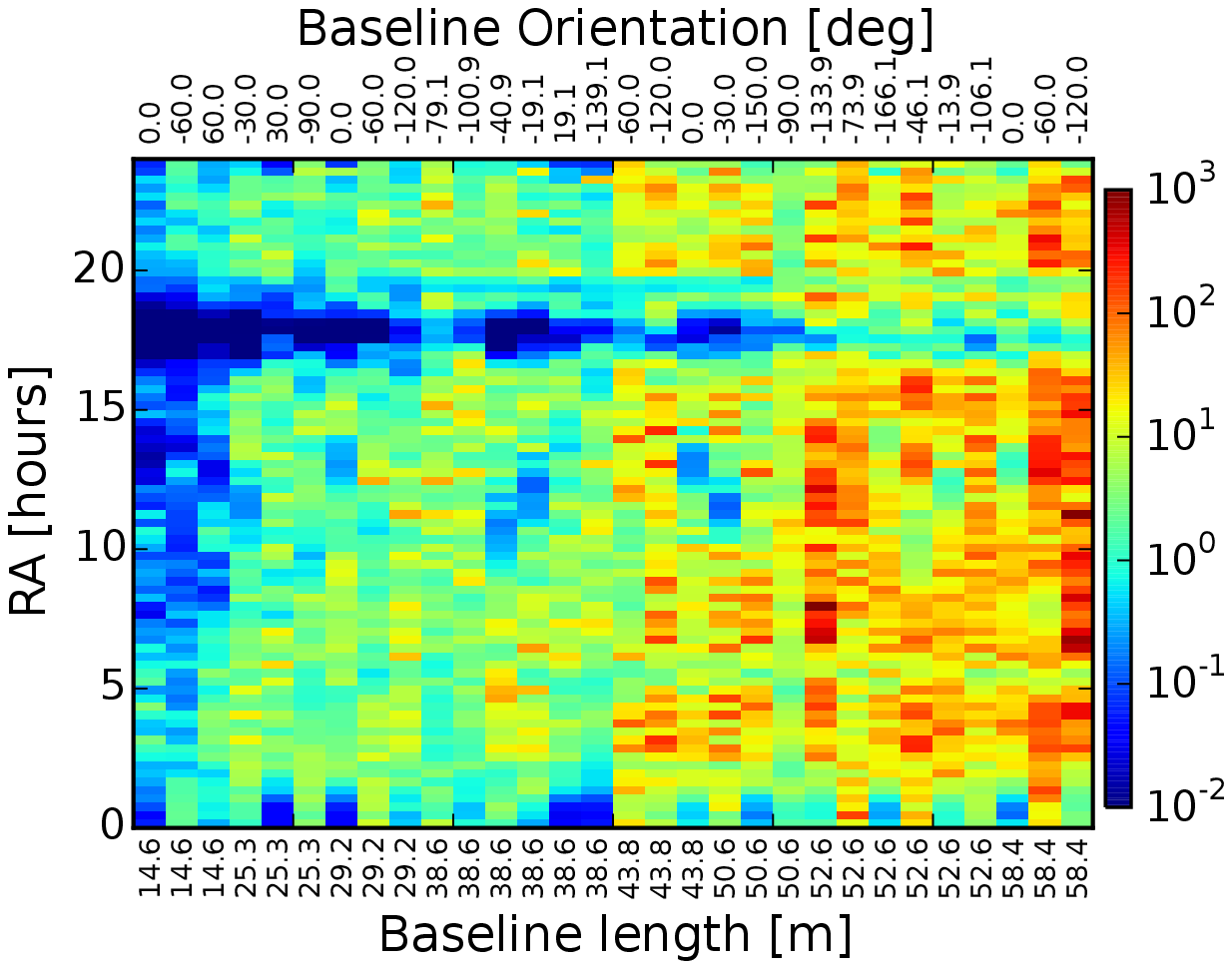}} \\
  \subfloat[][EoR model 1, 170~MHz subband ]{\label{fig:21cmfast-fg-ratio-170}\includegraphics[width=0.45\linewidth]{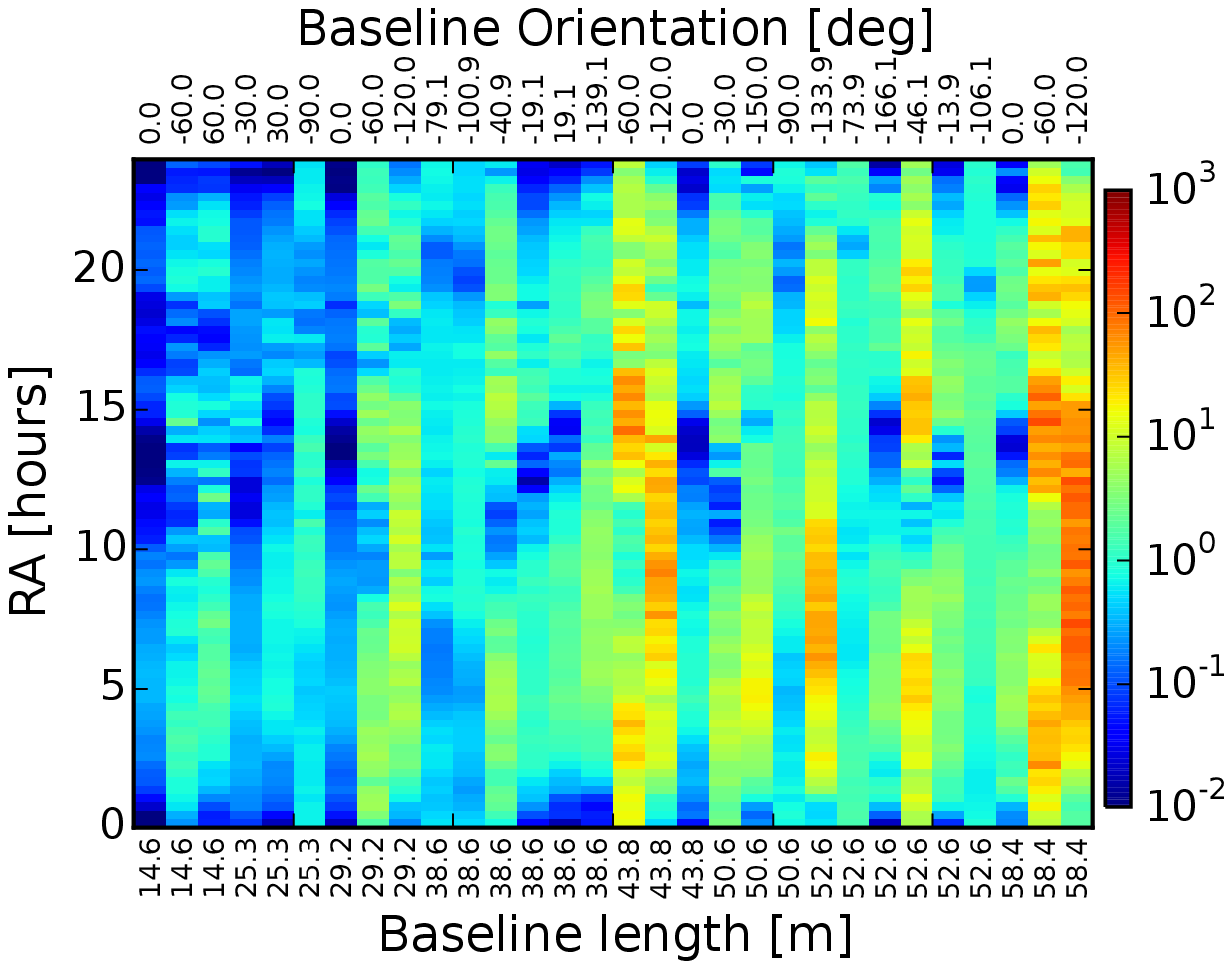}}
  \subfloat[][EoR model 2, 170~MHz subband ]{\label{fig:lidz-fg-ratio-170}\includegraphics[width=0.45\linewidth]{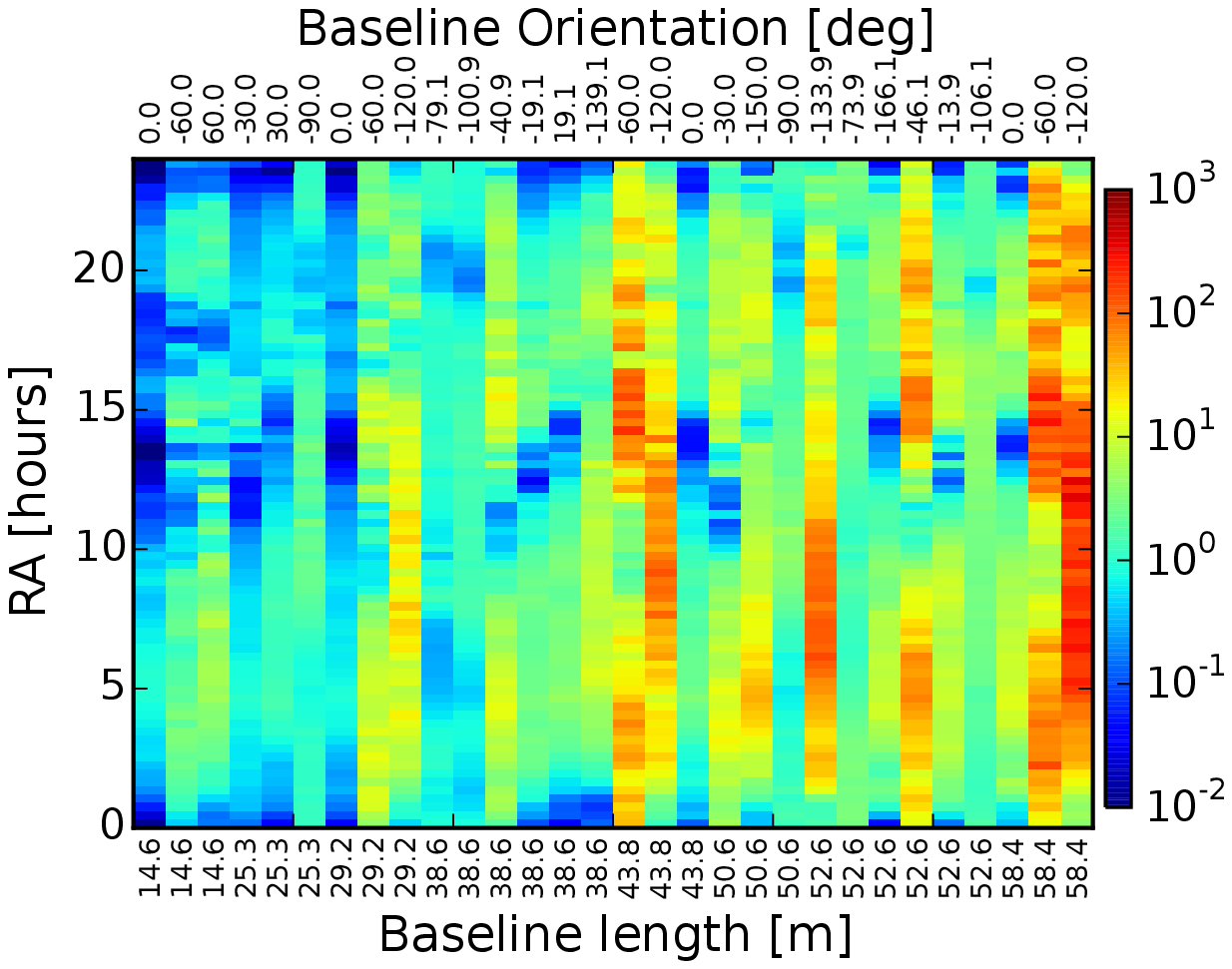}}
  \caption{Worst case EoR-foreground dynamic range, $\rho(\boldsymbol{b},f_0,\alpha)$, in {\it foreground avoidance} modes for different HERA-19 baselines as a function of RA ($\alpha$) of pointing on the sky. Each panel corresponds to the EoR model and subband indicated. The baseline vector is specified by its length (in m) and orientation (in degrees) at the bottom and top of the $x$-axis respectively. The color scale is logarithmic. Even in the worst case, HERA will be able detect EoR drawn from independent models on all baselines for a large fraction of a full 24~hour sky transit cycle. $\rho(\boldsymbol{b},f_0,\alpha)$ is seen to increase with increasing baseline length. This is because the loss in sensitivity to diffuse foreground power with increase in antenna spacings is faster than in the case of the EoR signal.}
  \label{fig:eor-fg-ratios}
\end{figure*}

We consider the foreground power spectrum obtained with the simulated chromatic antenna beam, and EoR models 1 and 2, in each of the 150~MHz and 170~MHz subbands. Fig.~\ref{fig:eor-fg-ratios} shows $\rho(\boldsymbol{b},f_0,\alpha)$ for all 30 unique HERA baselines over 24 hours of $\alpha$ for the two EoR models and two subbands used in this study. 

For the baselines considered, $\rho(\boldsymbol{b},f_0,\alpha)$ increases with baseline length. We attribute this to the much steeper dependence of foreground emission (predominantly diffuse emission) on baseline length than that of EoR H{\sc i} power. This is also confirmed from Fig.~\ref{fig:subbands} where the peak and sidelobe levels of foreground power obtained with the simulated chromatic antenna beam on the 14.6~m baseline, progressively drops by two orders of magnitude on the 58.4~m baseline. Hence, from a foreground contamination standpoint, the shortest baselines are less sensitive than longer ones in HERA-19 layout.

However, in all four cases considered here, except when the Galactic center appears overhead in the RA range roughly in the range 16-19 hr, the worst case signal-foreground ratio lies above unity in most of this parameter space. It demonstrates that HERA should detect the EoR signal with a very high signal-foreground ratio on all baselines over a majority of the pointings. 

\section{Summary}\label{sec:summary}

First-generation EoR experiments such as the MWA, PAPER and LOFAR have made enormous progress in the field of 21~cm cosmology. However, they are currently limited by the systematics due to extremely bright foregrounds and the instrument. Hence, characterizing these to a very high level of precision is critical for the success of next-generation EoR experiments. In this paper, we highlight the importance of a key systematic feature of the instrument -- the spectral structure of the antenna power pattern -- and its impact on EoR signal contamination, using HERA as an example. 

Using a new high dynamic range spectral weighting function obtained by convolving a {\it Blackman-Harris} window with itself, we isolate the effects of beam chromaticity on foreground delay power spectrum. We use three antenna beam models of varying chromaticity (achromatic, {\it Airy} and simulated chromatic) and show that increase in chromaticity extends the foreground spillover well beyond the horizon limits of the {\it foreground wedge} into the clean {\it EoR window} considered sensitive for EoR signal detection. The level of foreground spillover in the simulated chromatic beam is a few orders of magnitude worse than that from a nominal {\it Airy} disk pattern, which in turn is many orders of magnitude worse than that from an achromatic beam with no spectral structure. 

We note that inevitable reflections in the antenna-feed assembly and between reflecting structures across multiple antennas are also significant factors that contaminate the cosmological signal in $k_\parallel$-modes in the {\it EoR window}. Using a novel approach, we provide a formalism to set cosmologically-motivated design requirements directly on the level of suppression required on these reflections in the instrument. This provides a way to set goals on antenna performance and evaluate antenna designs that deliver larger collecting area per antenna.

By accounting for these instrumental systematics, if data is not limited by thermal noise, we demonstrate that HERA should detect EoR in $k_\parallel\gtrsim 0.2\,h$~Mpc$^{-1}$ with high significance with a simple spectral weighting even under a conservative analysis criterion that does not involve {\it foreground removal} strategies. All baselines of HERA-19 will not only detect but also help in filtering highly likely EoR models. 

We are also investigating advanced strategies for foreground removal that will robustly eliminate these systematics further. Advances in calibration, foreground modeling and subtraction will improve the performance listed here. More optimal analysis techniques and further design improvements will make HERA not only clear the obstacles limiting first-generation experiments but also become transformational for next generation of low-frequency cosmology experiments.

\acknowledgments

We thank Zaki Ali, Adam Beardsley, and Daniel Jacobs for their valuable inputs during the preparation of this manuscript. This work was supported by the U.S. National Science Foundation (NSF) through awards AST-1440343 and AST-1410719. ARP acknowledges support from NSF CAREER award 1352519. AEW acknowledges support from NSF Graduate Research Fellowship under grant 1122374.


\bibliographystyle{apj}

\end{document}